%% file: 1_MalIoT.tex
\documentclass{article}[10]
\usepackage{PACISEPaper} 

\usepackage{times} 
\usepackage{soul} 
\usepackage[hyphens]{url} 
\usepackage[hidelinks]{hyperref} 
\usepackage[utf8]{inputenc} 
\usepackage[small]{caption} 
\usepackage{mathtools}
\usepackage{array}
\usepackage{graphicx}
\usepackage{amsmath}
\usepackage{booktabs}
\usepackage{framed}
\usepackage{float}
\usepackage{relsize}
\usepackage[linesnumbered,ruled]{algorithm2e}
\usepackage{lipsum}
\usepackage{tkz-graph}
\usepackage{titlesec}
\usetikzlibrary{arrows}

\usepackage{etoolbox}

\begin{document}
%
\title{MalIoT: Scalable and Real-time Malware Traffic Detection for IoT Networks}
\author{
Ethan Weitkamp, Yusuke Satani, Adam Omundsen, Jingwen Wang, Peilong Li
\affiliations
Elizabethtown College, Computer Science Department
\emails
{
\href{mailto:weitkampe@etown.edu}{weitkampe}, 
\href{mailto:sataniy@etown.edu}{sataniy},
\href{mailto:omundsena@etown.edu}{omundsena},
\href{mailto:wangjingwen@etown.edu}{wangjingwen}, 
\href{mailto:lip@etown.edu}{lip}}
@etown.edu
}

\maketitle

\titleformat{\section}
  {\normalfont\fontsize{10}{0}\bfseries}{\thesection}{1em}{}
\setlength{\parskip}{0em}
\input{2_abstract}

\setlength{\parskip}{1em}

\titleformat{\section}
  {\normalfont\fontsize{12}{15}\bfseries}{\thesection}{1em}{}

\input{3_intro}

\input{4_related}
\input{5_design}
\input{6_evaluation}

\input{7_conclusion}
\input{8_ack}

\makeatletter
\renewenvironment{thebibliography}[1]
     {\section*{\refname}%
      \@mkboth{\MakeUppercase\refname}{\MakeUppercase\refname}%
      \list{\@biblabel{\@arabic\c@enumiv}}%
           {\settowidth\labelwidth{\@biblabel{#1}}%
            \setlength{\itemindent}{\dimexpr\labelwidth+\labelsep}
            \leftmargin\z@
            \@openbib@code
            \usecounter{enumiv}%
            \let\p@enumiv\@empty
            \renewcommand\theenumiv{\@arabic\c@enumiv}}%
      \sloppy
      \clubpenalty4000
      \@clubpenalty \clubpenalty
      \widowpenalty4000%
      \sfcode`\.\@m}
     {\def\@noitemerr
       {\@latex@warning{Empty `thebibliography' environment}}%
      \endlist}
\makeatother
\bibliographystyle{PACISE}
\bibliography{References}

\end{document}

%% file: 2_abstract.tex
\begin{abstract}

The machine learning approach is vital in Internet of Things (IoT) malware traffic detection due to its ability to keep pace with the ever-evolving nature of malware. Machine learning algorithms can quickly and accurately analyze the vast amount of data produced by IoT devices, allowing for the real-time identification of malicious network traffic. The system can handle the exponential growth of IoT devices thanks to the usage of distributed systems like Apache Kafka and Apache Spark, and Intel's oneAPI software stack accelerates model inference speed, making it a useful tool for real-time malware traffic detection. These technologies work together to create a system that can give scalable performance and high accuracy, making it a crucial tool for defending against cyber threats in smart communities and medical institutions.

\end{abstract}

%% file: 3_intro.tex
\section{Introduction}
\label{sec:intro}

Since the emergence of the Internet of Things (IoT), the problem of malware attacks on IoT devices has remained a constant problem. Technology improvements have caused the difficulty to increase over time, both in terms of the volume and the variety of the attacks \cite{gibert2020rise}. According to research, malicious attacks on IoT devices have significantly increased recently. Zscaler's research shows that during the pandemic in 2022, malware attacks on IoT devices linked to business networks have risen by 700\% \cite{zscaler}. The development of the internet, social networks, smartphones, and IoT devices have allowed bad actors to produce malware that is more advanced than ever before.

To protect against online attacks and maintain network stability, real-time and scalable malware detection solutions are required to identify and stop malware traffic in IoT networks. Due to its capacity to automatically detect and react to emerging malware threats in real-time, machine learning (ML) is a key strategy in IoT malware traffic detection \cite{jindal2019neurlux}, \cite{vinayakumar2019robust}, \cite{fang2020android}. ML models can swiftly scan this data to identify and prevent malware activity because IoT devices create enormous amounts of data that make it challenging to detect malware using conventional approaches. Because ML techniques are flexible and adaptable, it is possible to continuously upgrade the models to detect new varieties of malware as they appear. Additionally, ML models have the capacity to learn from enormous volumes of data and recognize patterns in the data that can point to malware activity, enabling them to detect malware of previously unidentified forms. 

However, current machine learning-based solutions for detecting malware in network traffic often struggle to handle the growing number of IoT devices and detect malicious traffic with low latency. To address these issues, this paper proposes a scalable end-to-end network traffic analysis system that can detect malware in real-time. The system utilizes distributed systems such as Apache Kafka and Apache Spark, allowing for efficient scalability as the number of IoT devices grows. Furthermore, the use of Intel's oneAPI software stack for both machine learning and deep learning models has been shown to improve the inference speed of the model three-fold. Specifically, there are three main contributions to this paper. (1) We seek to overcome the class imbalance and model over-fitting issues from the prior works by enriching the ``IoT-23'' dataset and adding a more diversified dataset named ``ToN\_IoT''. (2) We accelerate the malware traffic detection inference speed with the use of Intel's oneAPI software. (3) We boost the system scalability and responsiveness by implementing a big data platform that includes Apache Kafka as the streaming engine and Apache Spark as the data processor. 

The organization of this paper is as follows. Section \S \ref{sec:related} summarizes the previous works on IoT malware detection. Section \S \ref{sec:design} depicts the overall design of the data pipeline. Section \S \ref{sec:eval} evaluates the performance of the methods. Lastly, Section \S \ref{sec:concl} concludes the paper.

%% file: 4_related.tex
\section{Related Works}
\label{sec:related}

One of the traditional approaches for detecting malware traffic is through the use of signature-based detection methods \cite{Hajiheidari:2019}, \cite{masdari2020survey}, which use a database of known malware signatures to identify and block traffic that matches these signatures. While this method can be effective in detecting known malware, one of its biggest limitations is its incapability to detect unknown attacks, and it does not cover the detection of large-scale attacks. 

To solve this problem, network behavior-based detection is proposed for malware traffic detection \cite{Nari2013AutomatedMC}. Behavioral modeling methods focus on identifying and blocking malicious traffic based on its behavior rather than its signature \cite{aslan2021intelligent}, \cite{galal2016behavior}, \cite{liu2011behavior}. This approach is more effective at detecting new or unknown malware as it is not reliant on a database of known malware signatures. However, it can be more resource-intensive and may produce false positives. Most existing studies in this area have limited scope, focusing on specific malware types, such as Bots, or on particular attack types, such as DoS, and anomalies in specific protocols or network layers \cite{StoneGross2009YourBI}. Another limitation of traditional detection solutions is that the conventional solution depends on one network premises; this cannot detect attacks that originated at different network premises. Another problem is there is the case that devices might still operate for a long time, even after infection. 

Another technique used in malware traffic detection is sandboxing, which involves creating a virtual environment in which to analyze and test suspected malware \cite{miramirkhani2017spotless}, \cite{liu2022enhancing}. This allows analysts to safely observe the behavior of the malware and determine its potential threats.

Machine learning methods are introduced in the previous works for malware traffic detection \cite{Barut:2020}. Kaluphahana et al. \cite{sudheera2021adept} proposed Adept, a security framework that detects bots attack and classifies them into attack stages across space and time. Adept utilizes alert-level and pattern-level information to classify the type and stages of attacks into categorical classification by using machine learning models, k-nearest neighbor, random forest, and support vector machine. Moreover, an end-to-end monitoring system RTC was proposed to identify new threats by manually extracting features from different protocols and network layer traffic data \cite{Bekerman2015UnknownMD}. However, machine learning methods rely heavily on manual feature extraction, which is a time-consuming and resource-intensive process.

Since deep learning methods are efficient in feature extractions and automatic learning, the research on malware traffic attack detection is shifting from machine learning to deep learning \cite{Barut:2021}, \cite{Barut:2022}, \cite{Barut:2022:TNSM}, \cite{Huoh:2022}. Sahu et al. \cite{Sahu2021} present a security framework for IoT attack detection using a hybrid Deep Learning model with two stages. Specifically, a CNN model first learns the features from the IoT network traffic, then the feature representation generated from the previous step is used as the input of an LSTM model for attack detection. However, whenever the network is scaled up by adding additional IoT devices, then an additional CNN module should be accompanied by the connected network switch.

Table \ref{tbl:accuracy_survey} shows the performance of machine learning and deep learning models in classifying large-scale offensive accesses within IoT systems. Goyal et al. \cite{goyal2019http} observe that SVM and ANN models generate similar accuracy scores, while the precision score of ANN model outperforms SVM by 4.3\%. Long-Short-Term Memory (LSTM) is used to determine whether an attack or a benign attack can be detected from a relatively small amount of data. Liang et al. \cite{liang2019long} conclude that the SVM model has an accuracy score of 88.18\%, but the recall and F1 scores are below 50. On the other hand, the LSTM model has an accuracy score of 99.98\%, and both recall and F1 are close to 100.

\begin{table}[]
\scriptsize
\centering
\caption{Model Accuracy in Surveys}
\begin{tabular}{|l|l|l|l|l|l|}
\hline 
\textbf{Reference} & \textbf{Model} & \textbf{Accuracy}    & \textbf{Precision}    & \textbf{Recall}    & \textbf{F1}    \\ \hline \hline
Goyal et al. \cite{goyal2019http}    & ANN        & 99.74 & 95.99 & 100   & 97.95 \\ \hline
      & SVM        & 99.86 & 91.98 & 100   & 95.82 \\ \hline \hline
Liang et al. \cite{liang2019long}   & LSTM       &   -    & 99.98 & 100   & 99.99 \\ \hline
      & SVM        &    -   & 88.18 & 45.43 & 59.97 \\ \hline
\end{tabular}
\label{tbl:accuracy_survey}
\end{table}

%% file: 5_design.tex
\section{Design}
\label{sec:design}

In this section, we present the design of the MalIoT system by describing the overall data pipeline, data collection and generation, offline model training, and online model inference.

\subsection{Overview to the Data Pipeline}

\begin{figure}[!ht]
	\centering
	\includegraphics[width=3.4in]{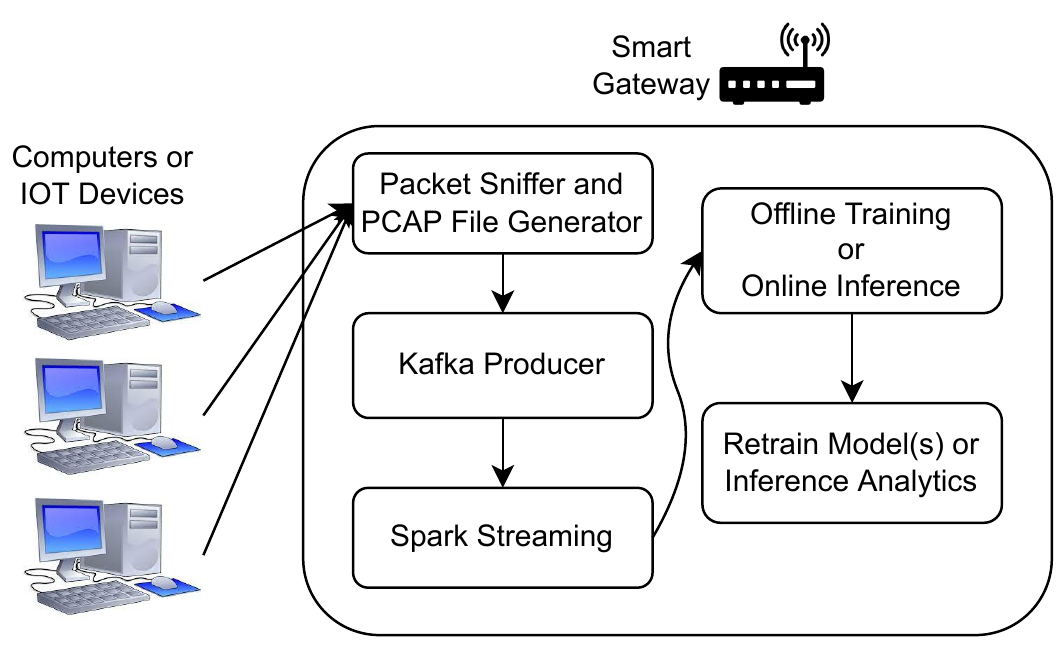}
	\caption{Overall Design Pipeline}
	\label{figure:design_pipeline}
\end{figure}

The overall data pipeline is illustrated in Fig. \ref{figure:design_pipeline}.
The information originates from IoT devices within a local network, which transmits their network activity through a smart gateway that houses the subsequent steps in the process. Initially, the network activity is intercepted by a software sniffer Bro on the smart gateway, which generates PCAP files from the received data. These PCAP files are subsequently dispatched to a Kafka producer, which transmits the data to a Kafka topic. The Kafka producer ensures that the pipeline can be scaled to accommodate additional IoT devices on the network. The data is then ingested from the Kafka topic using Spark streaming, which maintains the pipeline's scalability for any number of IoT devices. The data can then be utilized for the purposes of retraining ML or DL models, or for conducting an online inference on the network activity using previously trained ML or DL models.

\subsection{Data Generation}
\begin{figure}
	\centering
	\includegraphics[width=3.3in]{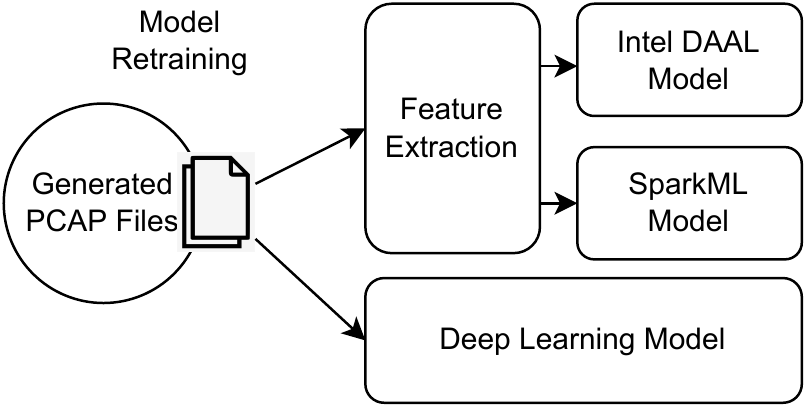}
	\caption{Model Training Process}
	\label{figure:model_retrain}
\end{figure}

To train various models, we need an initial dataset and a method to retrain the models using fresh data to adjust to new types of attacks. The retraining process is illustrated in Figure \ref{figure:model_retrain}. When PCAP files are produced from the packet sniffer, they can subsequently be used to retrain the models with the real-time network traffic on the network. We employed the IoT-23 Dataset and TON\_IoT Dataset as our starting datasets for training the original models. We tested the accuracy of machine learning and deep learning models using various feature subsets provided by the dataset and opted to use most of the features, excluding the host and recipient IP addresses and ports. These attributes didn't boost model accuracy, as they should not be linked with a malicious or benign attack and would merely overfit our models. Once the foundational models are established using the existing dataset, they can be retrained as needed using the newly generated PCAP files. For machine learning models, the desired features must be extracted from the PCAP files before retraining, as well as any other data preprocessing procedures.

\textbf{Datasets}: Aposemat IoT-23 datasets \cite{iot23} are captured from the network traffic of Internet of Things(IoT) devices. The dataset was first released in January 2020 and covers captures from 2018 to 2019. The IoT-23 datasets include 20 instances of malware captured from IoT devices, along with 3 instances of benign IoT device traffic. 

ToN\_IoT datasets \cite{Moustafa2021AND} are new generations of Internet of Things (IoT) and Industrial IoT (IIoT) datasets captured from a realistic and large-scale network with heterogeneous data sources. It consists of datasets from IoT and IIoT sensors, Operating systems datasets of Windows 7 and 10, Operating systems datasets of Ubuntu 14 and 18 TLS, and Network traffic datasets. 

The basic information of each dataset is listed in Table \ref{tbl:datasets}. IoT-23 datasets consist of 325,307,990 captures from different IoT network traffics, including 294,449,255 malware captures executed in infected IoT devices and 30,858,735 captures from benign IoT device traffic. In the ToN\_IoT datasets, the Windows 7 dataset has 28366 records and 132 features, the Windows 10 dataset consists of 35,975 records and 124 features, and the Network dataset has 21,978,632 records and 42 features. To generate the final synthesized training dataset, we integrate TON\_IoT and IoT-23 together and then supplement the synthesized dataset with extra benign traffic from the CIC IoT datasets \cite{Dadkhah2022TowardsTD} to balance the classification categories. 

\begin{table}[h!]
\scriptsize
\centering
\caption{Dataset Information}
\begin{tabular}{|c|c|c|}
  \hline
  \textbf{Dataset}  & \textbf{IoT-23} & \textbf{ToN\_IoT} \\
  \hline
  Benign  & 30,858,735 & 300,000\\
  \hline
  Anomaly  & 294,449,255 & 161,043\\
  \hline
  Total  & 325,307,990 & 461.043 \\
  \hline 
  \# of features & 23 & 45 \\
  \hline
  \# of IoT devices & 3 & 9 \\
  \hline
\end{tabular}
\label{tbl:datasets}
\end{table}

\textbf{Features}:
The common features shared in between the two datasets are listed in Table \ref{tbl:datasets_features}. In the experiments, we utilized two types of feature selection for our machine learning models: 1) the ``Full Feature Set'', which contains all features except for timestamp and unique identifier, and 2) the ``De-identified Feature Set'', which removes the TCP/UDP IP address and port information of both the originating and responding endpoints. We removed this information to avoid potential biases in the machine learning models when classifying malicious traffic based on IP addresses and ports.

\begin{table}[h!]
\scriptsize
\centering
\caption{IoT-23 and TON\_IoT Features Selection}
\begin{tabular}{|c|c|c|}
   \hline
   \textbf{IoT-23 Feature}  & \textbf{TON\_IoT Feature} & \textbf{Description} \\
   \hline
   ts & ts & Timestamp of connection  \\
   \hline
   id.orig\_h & src\_ip & originator's IP address \\
   \hline
   id.orig\_p & src\_port & originator's TCP/UDP port \\
   \hline
   id.resp\_h & dst\_ip & responder's IP address \\
   \hline
   id.resp\_p & dst\_port & responder's TCP/UDP port \\
   \hline
   proto & proto & transport layer protocol of connection \\
   \hline
   service & service & service \\
  \hline
   duration & duration & duration \\
   \hline
   orig\_bytes & src\_bytes & originator's payload bytes \\
   \hline
   resp\_bytes & dst\_bytes & responder's payload bytes \\
   \hline
   conn\_state & conn\_state & connection state\\
   \hline
   missed\_bytes & missed\_bytes & missed\_bytes \\
   \hline
   orig\_pkts & src\_pkts & number of ORIG packets \\
   \hline
   orig\_ip\_bytes & src\_ip\_bytes & number of ORIG IP bytes \\
   \hline
   resp\_pkts & dst\_pkts & number of RESP packets \\
   \hline
   resp\_ip\_bytes & dst\_ip\_bytes & number of RESP IP bytes \\
   \hline
 \end{tabular}
 \label{tbl:datasets_features}
 \end{table}

\subsection{Offline Model Training}
As time progresses, new malicious attacks emerge and are designed. What was once considered secure must adapt to maintain its security in the present and future. This is especially true in the field of machine learning. To ensure our predictive models remain secure, they must be trained on more up-to-date data. To support this, we have integrated a means of retraining our various models with our own generated data. The data passing through the pipeline is saved in a format that enables easy retraining of our models at a later point in time. Therefore, we need not rely solely on existing datasets to keep our models up to date. 

We evaluated five machine learning models and five deep learning models. The machine learning model classifiers include random forest, decision tree, logistic regression, linear SVC, and Gaussian-NB. The deep learning models consist of artificial neural networks (ANN), 1D convolutional neural networks (1DCNN), two-dimensional CNN (2DCNN), long short-term memory (LSTM), and a combination of CNN and LSTM. Each model possesses unique strengths and benefits that prompted our decision to evaluate them. Regarding the hyperparameters of our deep learning models, we used ReLU as the activation function for all of the ANN and CNN models, except for the LSTM model, which uses `tanh` as the activation function and sigmoid as the recurrent activation function. The ANN only uses one layer, while all of the other models use two layers. More details about our model hyperparameters can be found in Table \ref{tbl:parameters}.
\begin{table}[h!]
\scriptsize
\centering
\caption{Deep Learning Hyper-Parameters}
\begin{tabular}{|c|c|}
   \hline
   \textbf{Parameter}  & \textbf{Value} \\
   \hline
   learning\_rate &  1e-3\\
   \hline
   decay\_rate & 1e-5\\
   \hline
   dropout\_rate & 0.5\\
   \hline
   dense\_units & 128\\
   \hline
   n\_batch &  100\\
   \hline
   n\_epoch & 1\\
   \hline
   filters & filters\\
  \hline
   kernel\_size & 4\\
   \hline
   strides & 1\\
   \hline
   CNN\_layers & 2\\
   \hline
   clf\_reg & 1e-5\\
   \hline
 \end{tabular}
 \label{tbl:parameters}
 \end{table}
 
\subsection{Online Inference}

After the network traffic is converted into PCAP files and processed by Spark streaming, our previously trained model(s) can make an inference on the data. As mentioned previously, we have trained five types of machine learning models and five types of deep learning models.

Upon entering the Spark streaming stage of the pipeline, all new data is evaluated by one of the trained models based on the data's timestamp. Machine learning models generally require longer preprocessing times since feature extraction is necessary, but inference/prediction times are quick as these models are not very complex. On the other hand, deep learning models have less preprocessing time but longer inference times due to their complexity. These models tend to be more accurate on complex data, like that found in our pipeline. At present, the pipeline concludes once an inference is made on the network traffic, but the resulting inferences may be used as desired.

%% file: 6_evaluation.tex
\section{Evaluation}
\label{sec:eval}
This section of the paper evaluates the performance of our proposed system. Specifically, we will talk about the specifications of the testbed, and the accuracy and timing performance of each machine learning and deep learning model.

\subsection{Experiment Testbed}
We summarize the system specifications and software versions for the testbed in Table \ref{tbl:platform}. We use built-in Python tools in order to collect our data, like training time, inference time, accuracy, and CPU usage.

\begin{table}[h!]
\scriptsize
\centering
\caption{Experiment Platform Specification}
\begin{tabular}{|c|c|}
  \hline
  \textbf{Item}  & \textbf{Specification} \\
  \hline
  CPU &  Intel(R) Core(TM) i9-9920X CPU @ 3.50GHz\\
   \hline
  GPU & GTX 2080 Ti with 11 GB DDR6 Memory\\
  \hline
  Memory & 64 GB DDR4 @ 3600MHz\\
  \hline
  Storage & 2TB SSD\\
  \hline
  Host OS & Ubuntu 18.04 LTS\\
  \hline
  Tensorflow & 2.7 \\
  \hline
  Apache Spark & 3.0.1 \\
  \hline
  Apache Kafka & 2.6.0 \\
  \hline
  Accelerator & Intel DAAL v2020.1\\
  \hline
\end{tabular}
\label{tbl:platform}
\end{table}

\subsection{Model Accuracy Comparison}
We calculated the prediction accuracy from machine learning and deep learning models. In the machine learning inference section, we design five machine learning models (Random-forest, Decision-Trees, Logistic-Regression, Linear-SVC, and Gaussian-NB) and two feature sets, the full feature set, and the de-identified feature set. 

Figure 3 shows the outcomes of machine learning models. In addition, Figure 4 shows the outcomes from five deep learning models, Artificial Neural Network, Convolutional Neural Network, Convolutional Neural Network\_2D, Long-Short-Term-Memory, and a combination of Convolutional Neural Network and Long-Short-Term-Memory.

 \begin{figure}[h!]
	\centering
	\includegraphics[width=3.3in]{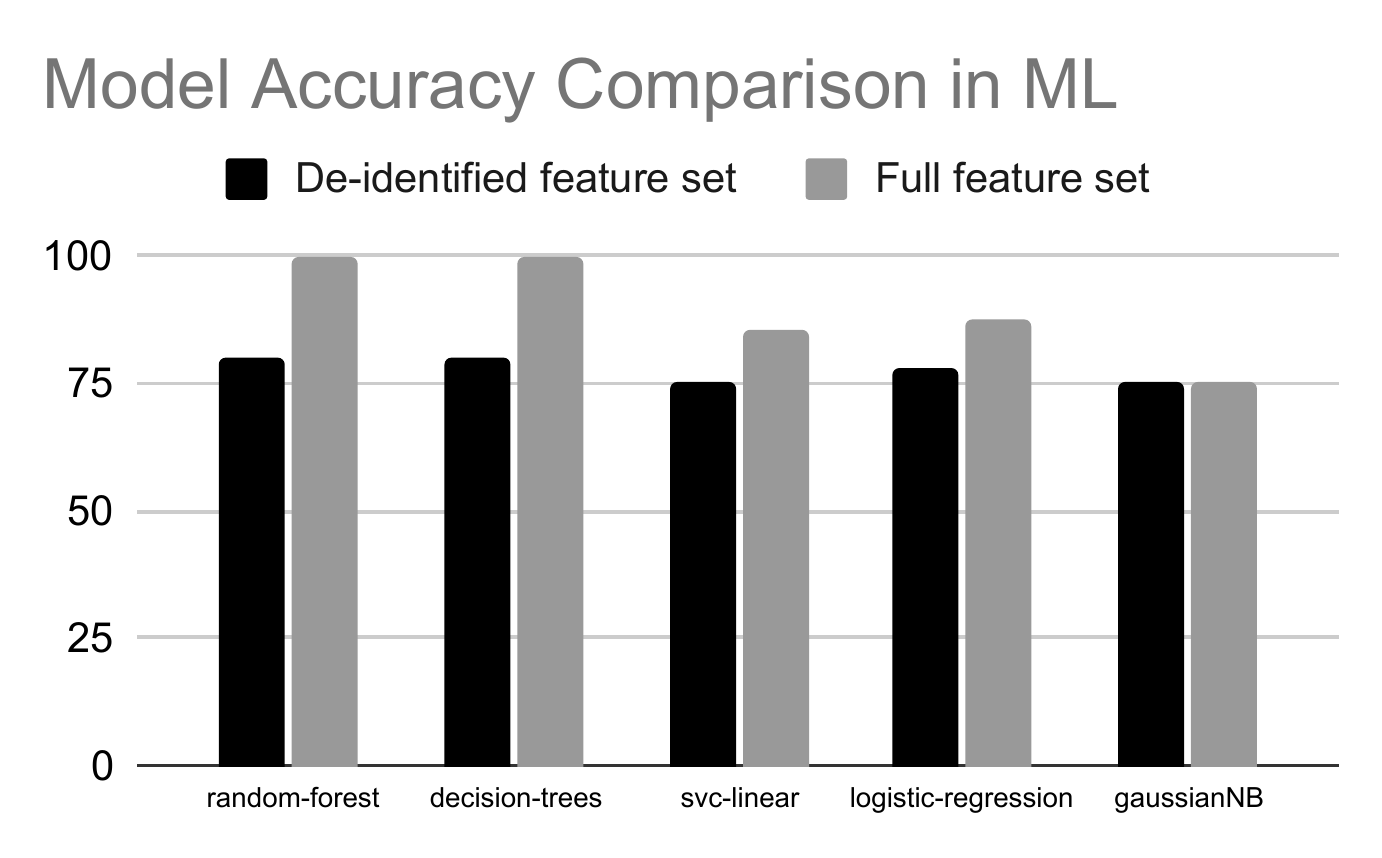}
	\caption{Model Accuracy Comparison in ML}
	\label{figure:training-time-netml}
\end{figure}

In the machine learning experiments, generally full feature set marked a higher score. The first two models, random-forest and decision-trees, almost reached 100\%. The score of the de-identified feature set is 75\% on the whole.

 \begin{figure}[h!]
	\centering
	\includegraphics[width=3.3in]{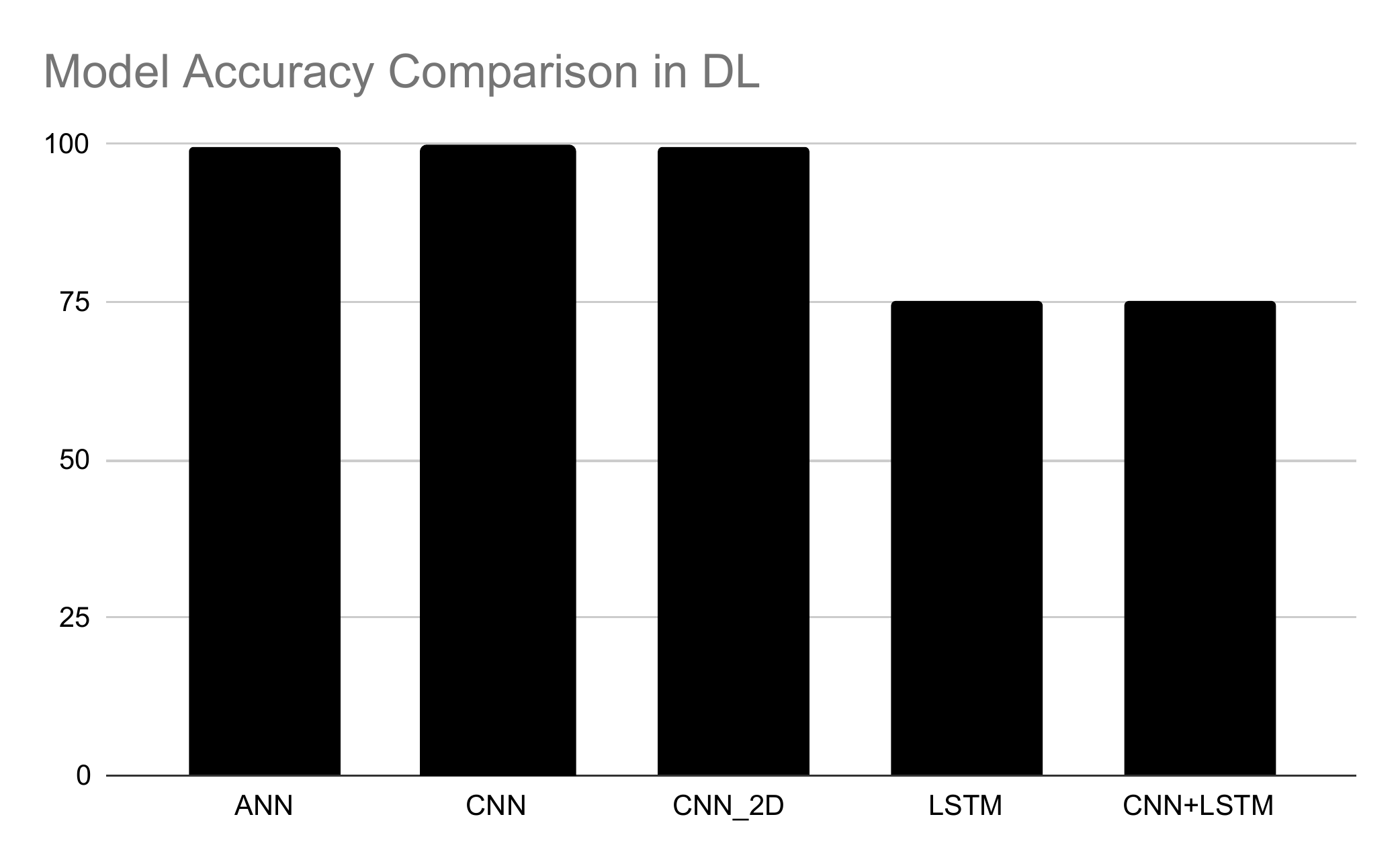}
	\caption{Model Accuracy Comparison in DL}
	\label{figure:training-time-netml}
\end{figure}

In the deep learning section, the first three models were marked as almost 100\%. LSTM and LSTM+CNN, on the other hand, scored about 75\%.

\subsection{Inference Time Comparison}
\begin{figure}[h!]
	\centering
	\includegraphics[width=3.3in]{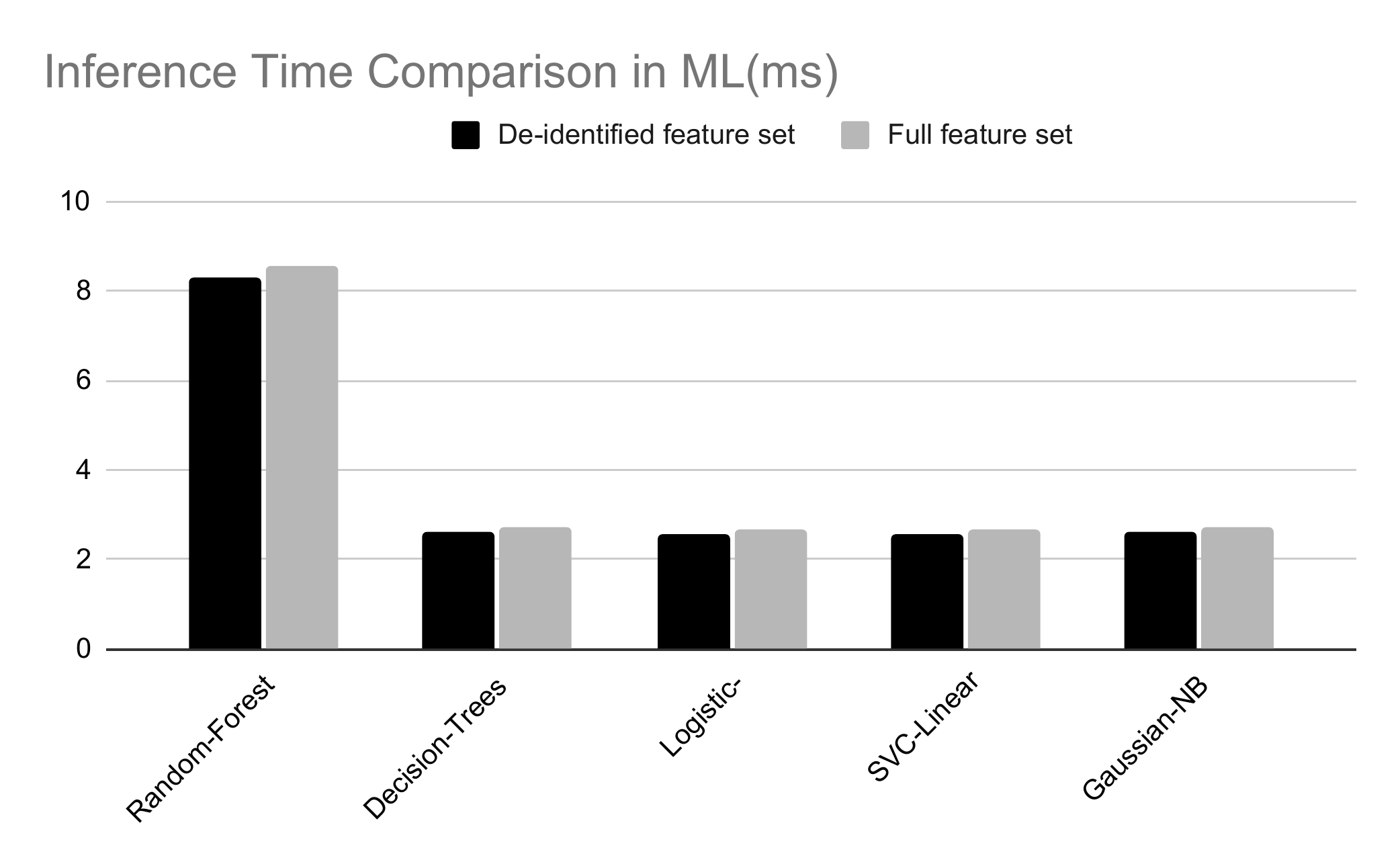}
	\caption{Inference Time Comparison in ML (ms)}
	\label{figure:inf-time-ml}
\end{figure}

\begin{figure}[h!]
	\centering
	\includegraphics[width=3.3in]{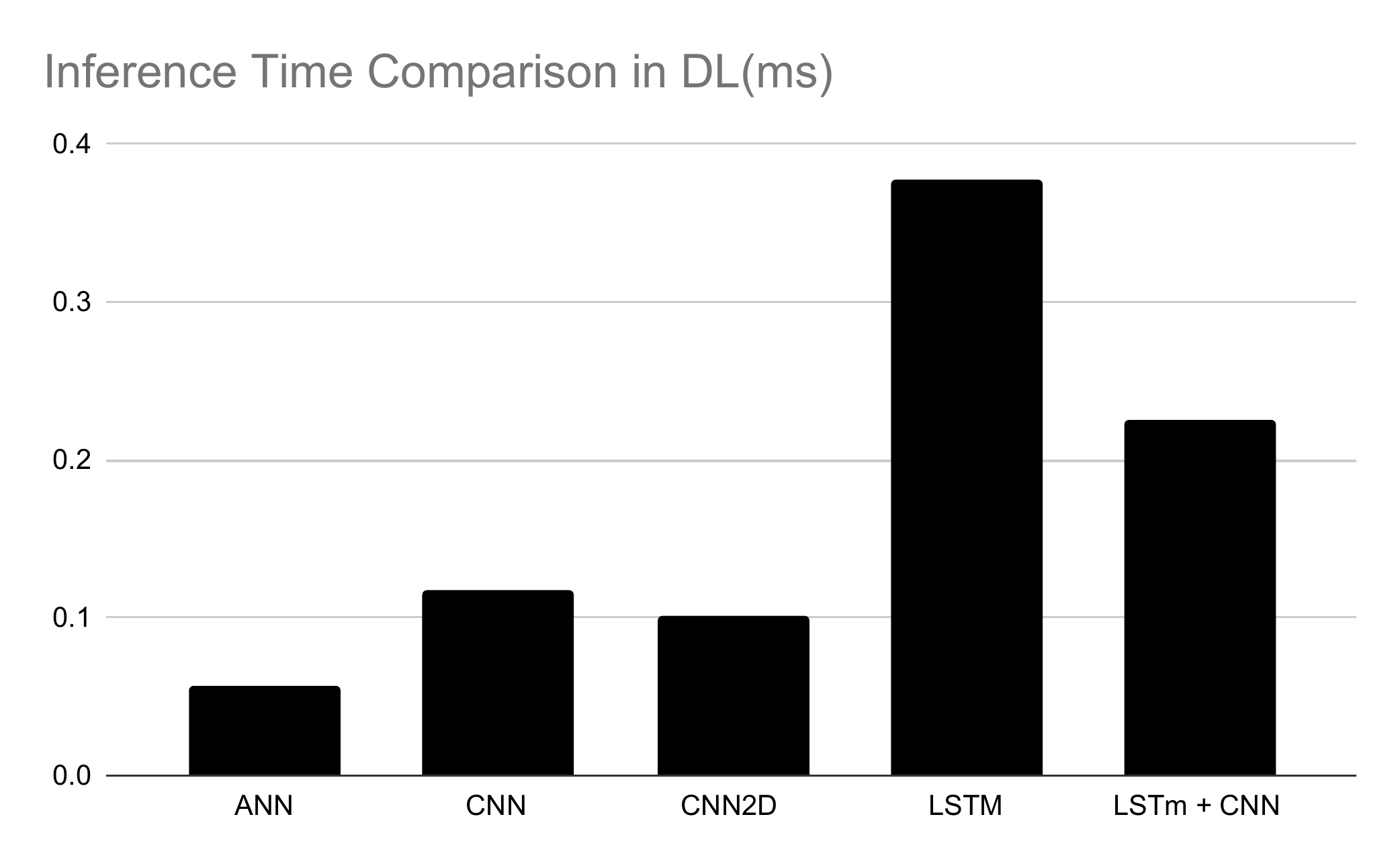}
	\caption{Inference Time Comparison in DL (ms)}
	\label{figure:inf-time-dl}
\end{figure}

We measured the time of average time it takes for inference per each line from CSV files. As same with before experiment, we tested machine learning and deep learning. Figure \ref{figure:inf-time-ml} Shows the results from machine learning models, and Figure \ref{figure:inf-time-dl} Shows the results from deep learning models

\begin{table}[]
\scriptsize
\centering
\caption{Inference Time for Machine Learning Models on Two Different Feature Sets}
\begin{tabular}{|c|c|c|}
\hline
                    & \textbf{De-identified feature set (ms)} & \textbf{Full feature set (ms)} \\ \hline
Random-Forest       & 33.18         & 34.20 \\ \hline
Decision-Trees      & 10.40           & 10.78 \\ \hline
Logistic-Regression & 10.30          & 10.71 \\ \hline
    SVC-Linear          & 10.30          & 10.68 \\ \hline
Gaussian-NB         & 10.48          & 10.87 \\ \hline
\end{tabular}
\label{tbl:inference_ml}
\end{table}


\begin{table}[]
\scriptsize
\centering
\caption{Inference Time for Deep Learning Models}
\begin{tabular}{|c|c|}
\hline
         & \textbf{Time per row (ms)} \\ \hline
ANN      & 0.056                  \\ \hline
CNN      & 0.118                   \\ \hline
CNN2D    & 0.102                   \\ \hline
LSTM     & 0.378                  \\ \hline
CNN+LSTM & 0.226                   \\ \hline
\end{tabular}
\end{table}

This experiment shows that the random-forest model takes longer than othez1r models. The reason seems that random-forest runs decision trees in parallel. The other models make inference per line around 2.6ms. In addition, the full feature set requires a little longer time to predict than the De-identified feature set.

This experiment shows that all deep learning models make inferences much faster than machine learning models. LSTM models take a little longer than other models, but only 0.4 ms per line

 \subsection{Scalability}
\begin{figure}[h!]
	\centering
	\includegraphics[width=3.3in]{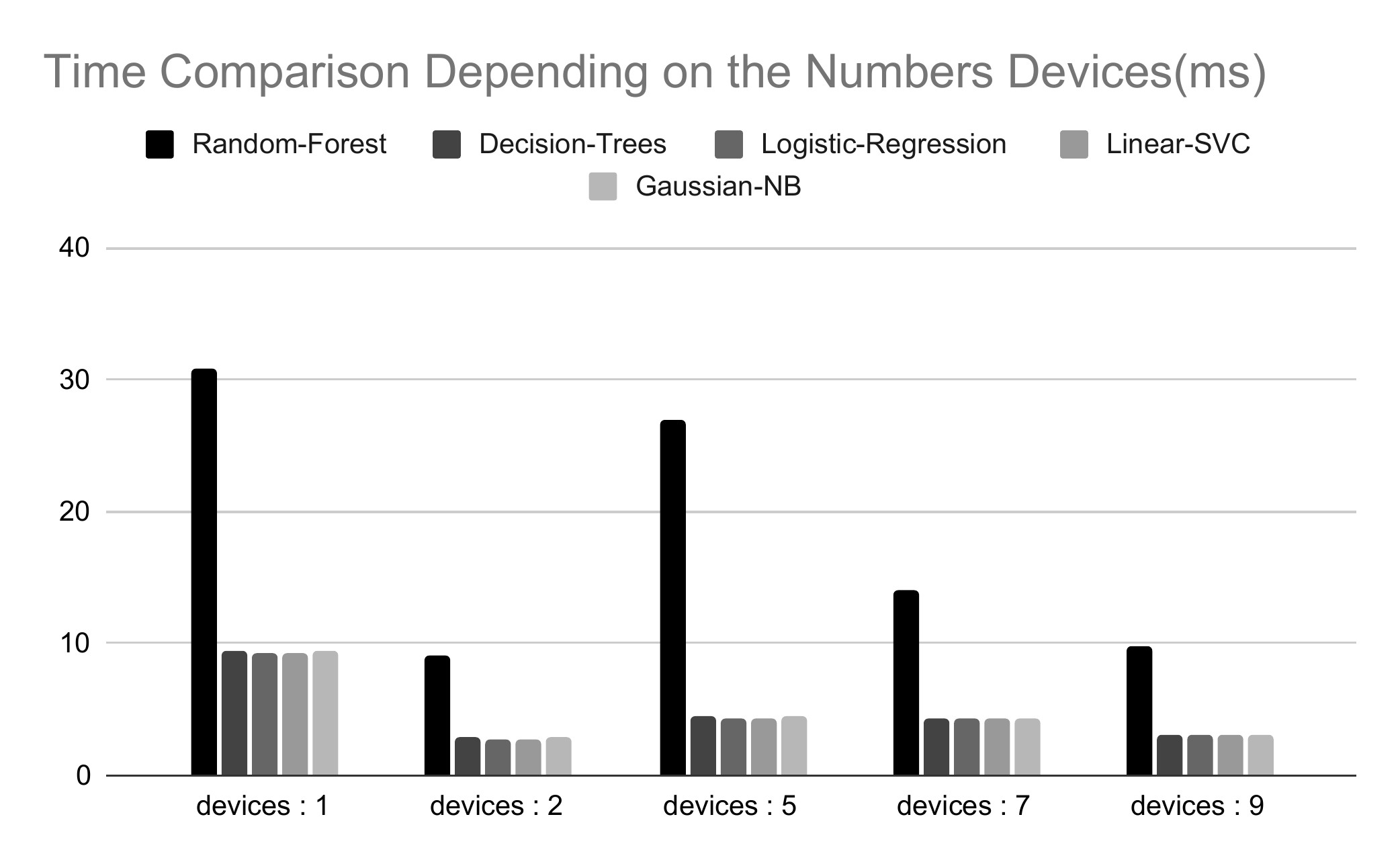}
	\caption{Time Comparison Depending on the Numbers Devices(ms)}
	\label{figure:scability_results}
\end{figure}

 Another key component of our research is scalability. We measured the change in inference time depending on the number of supporting devices.
In the test, we added the number of devices from one to nine. Figure \ref{figure:scability_results} Shows the outcome of the trial.

The experiment shows the scalability of our model. Generally, the inference time per line decreases as the number of connected devices gradually increases. Random-forest takes longer to predict than other models, the same as the previous experiment.

%% file: 7_conclusion.tex
\section{Conclusion}
\label{sec:concl}

In our project, we were able to achieve the objectives of real-time and scalable malicious network traffic detection by utilizing Apache Kafka producer and Apache Spark. Our models were trained on an enriched dataset that comprises IoT-23 and TON\_IoT which contains millions of network flows with information on both malicious and benign network traffic. Upon preprocessing the data, we were able to perform inferences on it using the various models we developed. This process could be automated on a smart gateway to enable real-time detection on a local network. We have made the source code for this project available on GitHub for future research purposes. It can be accessed via the following link: \url{https://github.com/BlueJayADAL/NetSec}.

%% file: 8_ack.tex
\section*{Acknowledgment}
This work is supported in part by a grant from Intel Corporation, and a grant from the Summer Scholarship, Creative Arts and Research Projects (SCARP) Program of Elizabethtown College.

%% file: 1_MalIoT.bbl
\begin{thebibliography}{10}
\bibitem{zscaler}
Zscaler study confirms iot devices are a major source of security compromise,
  reinforces need for zero trust security.
\end{thebibliography}
